\documentclass[12pt]{article}
\usepackage{hyperref}  
\usepackage{graphicx}
\usepackage{lineno}  

\newcommand\pubdate{\today}

\def\institute{Deutsches Elektronen-Synchrotron (DESY)\\
Notkestra{\ss}e 85, 22607 Hamburg, GERMANY}

\def\Title#1{\begin{center} {\Large #1 } \end{center}}
\def\Author#1{\begin{center}{ \sc #1} \end{center}}
\def\Address#1{\begin{center}{ \it #1} \end{center}}

\newcommand\pubblock{\rightline{\begin{tabular}{l}
         \pubdate  \end{tabular}}}	 
\newenvironment{Abstract}{\begin{quotation}  }{\end{quotation}}
\newenvironment{Presented}{\begin{quotation} \begin{center} 
             PRESENTED AT\end{center}\bigskip 
      \begin{center}\begin{large}}{\end{large}\end{center} \end{quotation}}

\def\beq{\begin{equation}}
\def\eeq#1{\label{#1}\end{equation}}
\def\eeqn{\end{equation}}

\def\beqa{\begin{eqnarray}}
\def\eeqa#1{\label{#1}\end{eqnarray}}
\def\eeqan{\end{eqnarray}}

\def\st{\scriptstyle}

\let\bar=\overbar

\def\Dslash{\not{\hbox{\kern-4pt $D$}}}
\def\dslash{\not{\hbox{\kern-2pt $\del$}}}

\def\msb{{\bar{\ssstyle M \kern -1pt S}}}

\def\pt{p_\mathrm{T}}
\def\ptmiss{\pt^\mathrm{miss}}
\def\st{S_\mathrm{T}}
\def\ttbar{\mathrm{t}\bar{\mathrm{t}}}
\def\TTbar{\mathrm{T}\bar{\mathrm{T}}}

\begin{document}
\begin{titlepage}
\pubblock

\vfill
\Title{Searches for vector-like quarks and resonances decaying to top quarks with the ATLAS and CMS detectors}
\vfill
\Author{ Gerrit Van Onsem \\ on behalf of the ATLAS and CMS experiments }
\Address{\institute}
\vfill
\begin{Abstract}
Many models beyond the standard model predict the existence of vector-like quarks or other types of heavy resonances. Using proton-proton collision data at center-of-mass energies of 8 and 13 TeV, the ATLAS and CMS Collaborations have performed a multitude of searches for pair and singly produced vector-like quarks decaying to W, Z, or Higgs bosons. Heavy gauge and (pseudo)scalar resonances decaying to top quarks have been searched for as well. No evidence for any of these hypothetical particles has been found yet, and stringent exclusion limits are derived on their masses, cross sections and couplings in various models. 
\end{Abstract}
\vfill
\begin{Presented}
$10^{th}$ International Workshop on Top Quark Physics\\
Braga, Portugal,  September 17--22, 2017
\end{Presented}
\vfill
\end{titlepage}
\def\thefootnote{\fnsymbol{footnote}}
\setcounter{footnote}{0}
%


\section{Introduction}
Vector-like quarks (VLQs) are hypothetical color-charged fermions whose left-handed and right-handed field components transform in the same way under the weak-isospin SU(2) gauge group, and are predicted in many models beyond the standard model (SM)~\cite{VLQhandbook}. 
Top-like and bottom-like VLQs, denoted by T and B, are assumed to have electric charges of magnitude $2/3$ and $1/3$ of the electron charge, respectively. Particles with more exotic charges of magnitude $5/3$ and $4/3$, denoted by X and Y respectively, are allowed as well. The T and B quarks may decay to a W, Z, or Higgs boson and a third-generation quark, but X and Y quarks are only allowed to decay to a W boson, due to charge conservation. Alternatively, VLQs may also couple to light-flavor quarks, and in those scenarios they will be denoted by Q. Interestingly, the weak-interaction multiplets in which VLQs would reside (singlets, doublets, or triplets) influences the decay branching fractions. The wide range of possible VLQ types, decay modes, and production mechanisms defines a broad search program at the CERN LHC. 

Other new resonances are predicted in many models beyond the SM, and if these resonances are heavy enough, they may decay to top quarks. These include, for instance, new gauge bosons $\mathrm{W}^\prime \rightarrow \mathrm{t}\bar{\mathrm{b}}$, $\mathrm{Z}^\prime \rightarrow \ttbar$, $\mathrm{Z}^\prime \rightarrow \mathrm{T}\bar{\mathrm{t}}$, or new (pseudo)scalar bosons $\mathrm{H(A)} \rightarrow \ttbar$. Interesting phenomena may appear due to the quantum interference of such final states with SM processes, and have to be taken into account in the experimental searches at the LHC. 

Many searches for these new particles have been performed with the ATLAS~\cite{ATLAS} and CMS~\cite{CMS} detectors, and no evidence for their existence has been found yet. This report focuses on highlights of the experimental searches since the 2016 edition of the series of International Workshops on Top Quark Physics\footnote{The $9^{th}$ International Workshop on Top Quark Physics in Olomouc, Czech Republic, September 19--23, 2016.}. 

\section{Searches for vector-like quarks}

\begin{figure}[htb]
\centering
\includegraphics[width=0.47\textwidth]{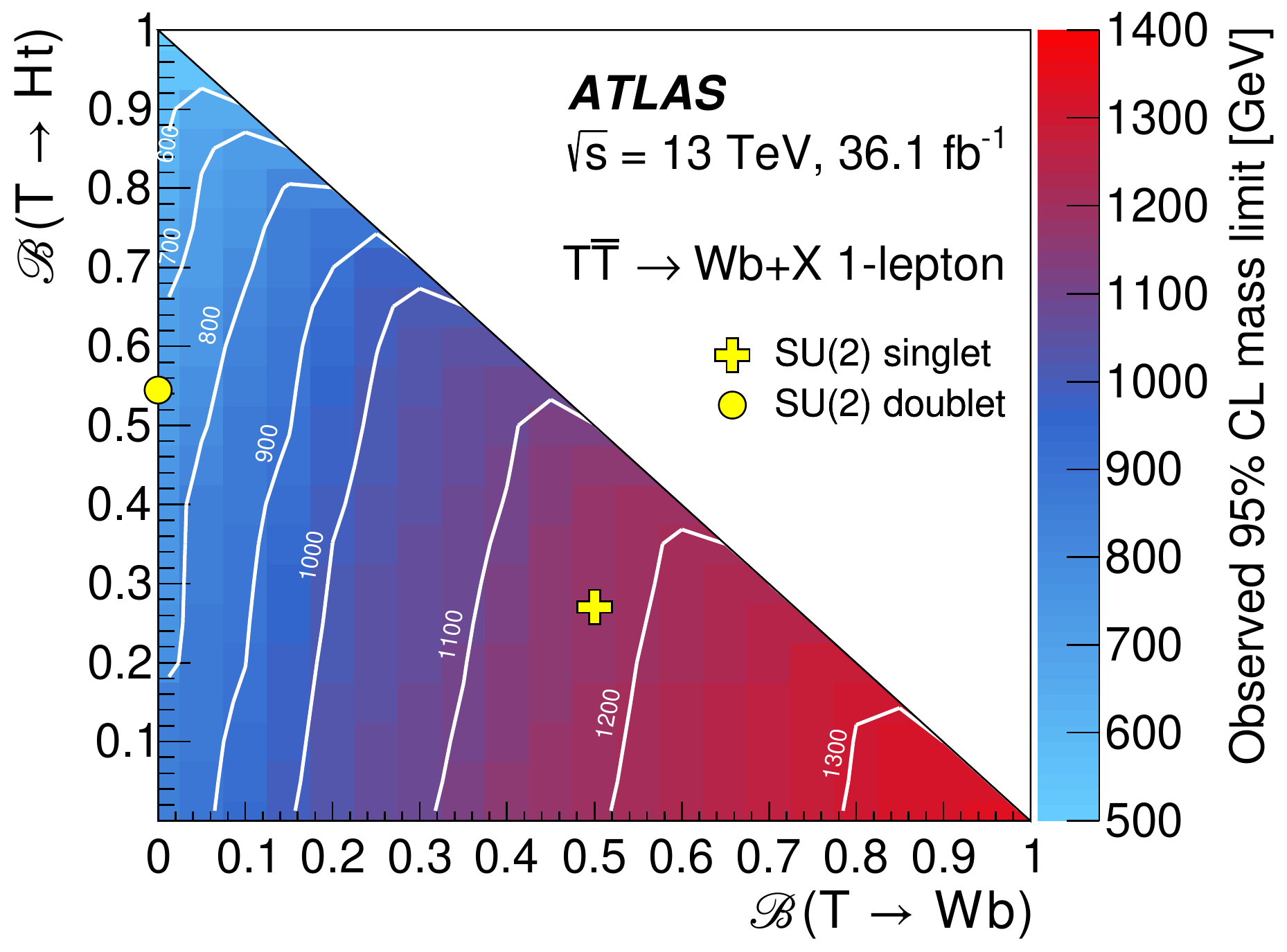}
\includegraphics[width=0.47\textwidth]{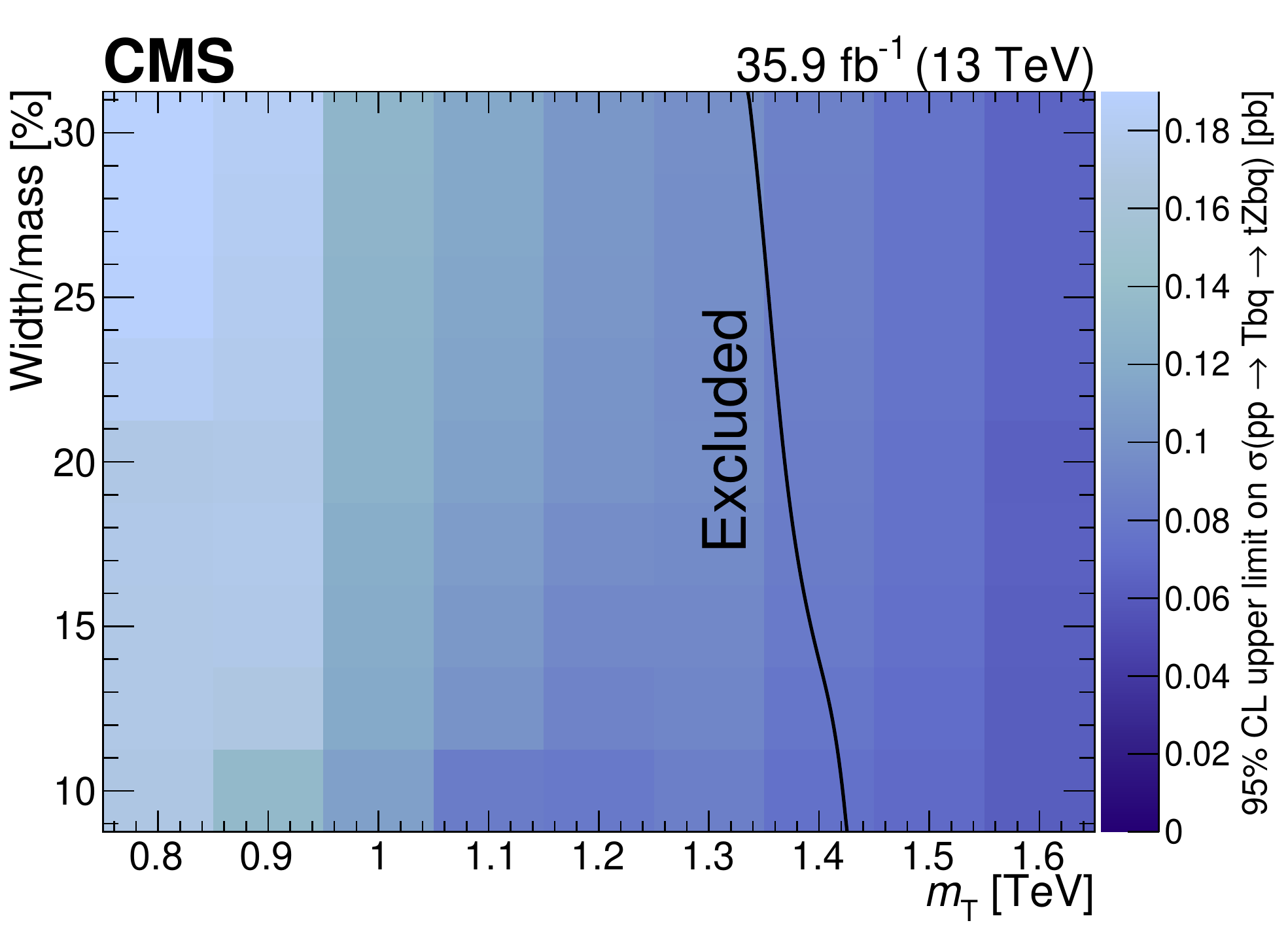}
\caption{Left: observed lower mass limits obtained in a search for pair produced T quarks ($\TTbar \rightarrow \mathrm{b}\mathrm{W} + \mathrm{x}$)~\cite{1707.03347}. Right: left-handed T cross section limits as a function of the decay width and mass, obtained in a search for singly produced T quarks ($\mathrm{T} \rightarrow \mathrm{t}\mathrm{Z}$)~\cite{1708.02510}.}
\label{fig:VLQlimits}
\end{figure}

A search has been performed with the ATLAS experiment for pair produced vector-like T quarks with a significant branching fraction to a W boson and a b quark~\cite{1707.03347}. The search includes an interpretation for vector-like B quarks as well.  
Events are selected with one isolated lepton, $\geq 3$ (small-radius) AK4 jets of which at least 1 b-tagged, $\geq 1$ (large-radius) AK10 W-tagged jets, and significant missing transverse momentum $\ptmiss$. A signal region and control region is defined, based on the angular separation $\Delta R(\ell,\nu)$ of the lepton and reconstructed neutrino, and $\st$, defined as the scalar sum of transverse momenta of objects in the event. The $\TTbar$ system is reconstructed, and the mass $m_\mathrm{T}^\mathrm{lep}$ of the leptonically decaying VLQ is used to search for the signal. Limits are derived via a binned maximum likelihood fit on $m_\mathrm{T}^\mathrm{lep}$, performed simultaneously in the signal and control regions. The resulting lower mass limit contours at 95\%~CL are calculated as a function of the VLQ decay branching fractions, for T quarks shown in Fig.~\ref{fig:VLQlimits} (left), and for B quarks. For 100\% branching fraction of the VLQ to a W boson as well as in the singlet scenarios, the lower limits are well above 1 TeV. 

The CMS Collaboration performed a search for pair produced vector-like T quarks purely decaying to W bosons~\cite{B2G-17-003}. The event selection requires a single isolated lepton, $\geq 4$ AK4 jets or 3 AK4 jets and 1 AK8 W-tagged jet, and $\st > 1000$~GeV. A kinematic fit is performed using the lepton, $\ptmiss$, AK4 jets not matched to AK8 jets, and subjets of W-tagged AK8 jets. The reconstruction hypothesis in this fit is the semileptonic decay of the $\TTbar$ pair. The used constraints are the W boson mass on the leptonic and hadronic side, and the equality of the reconstructed VLQ masses on both sides. Limits are derived using this reconstructed mass, and T masses below 1295 TeV are excluded. 

The ATLAS Collaboration searched for pair produced vector-like T quarks with a significant branching fraction to a Z boson~\cite{1705.10751}. As the search is optimized for the Z boson to decay to neutrinos, events are selected in the single lepton final state with large $\ptmiss$. 
The normalization of the $\ttbar$ and W+jets backgrounds are obtained via a fit in dedicated control regions, and the modeling is monitored in independent validation regions. A signal region is defined, based on the requirement of $\geq 2$ AK10 jets, $\ptmiss > 350$~GeV, and (generalized) transverse mass requirements suppressing $\ttbar$ and W+jets background events. 
Lower mass limits are deduced via simultaneous fits of the event yield in the signal and control regions, as a function of the VLQ branching fractions. For $\mathrm{T} \rightarrow \mathrm{t}\mathrm{Z}$ (100\%), the lower mass limit is 1.16~TeV, while for singlet and doublet scenarios the limits are 0.87~TeV and 1.05~TeV, respectively.

Two searches were performed with the CMS experiment for pair produced vector-like X quarks with exotic charge of magnitude 5/3. In the first search~\cite{B2G-16-019}, the same-sign dilepton final state is analyzed, with background events that include same-sign prompt leptons, opposite-sign prompt leptons, and same-sign non-prompt leptons. A counting experiment is performed after requiring that the scalar sum of $\pt$ of leptons and $\pt$ of jets in the event exceeds 1.2~TeV. Masses of right-handed (left-handed) X quarks are excluded below 1.16 (1.10)~TeV.
In the second search~\cite{B2G-17-008}, in the single lepton final state, a template fit is performed using the minimum mass among the lepton and b-tagged jets, in 16 categories based on lepton flavor, number of b-tagged jets, W-tagged jets, and top-tagged jets. Masses of right-handed (left-handed) X quarks are excluded below 1.32 (1.30)~TeV. 

The CMS Collaboration searched for singly produced T quarks decaying to a Z boson and a top quark~\cite{1708.01062}, taking into account a finite T width of up to 30\% of its mass. 
Events are selected with $\geq 1$ b-tagged jets and a dilepton pair compatible with the decay of a Z boson. Next, 10 categories are defined based on lepton flavor, the type of top quark candidates (either with fully merged, partially merged, or resolved decay products), and the number of forward jets, since an associated forward jet is characteristic of single production. The mass of the T quark is reconstructed from the top quark and Z boson candidates. Limits are derived from a fit using the reconstructed mass distributions in the 10 categories, as a function of the mass and the decay width of the T quark, as shown in Fig.~\ref{fig:VLQlimits} (right). This analysis also provides an interpretation for a $\mathrm{Z}^\prime$ resonance decaying to a vector-like T quark and a top quark, with excluded cross sections below $[0.13,0.06]$~pb for $\mathrm{Z}^\prime$ and T masses in the range of $[1.5,2.5]$ and $[0.7,1.5]$~TeV, respectively.

A search for singly produced vector-like B quarks decaying predominantly to a bottom quark and a Higgs boson was also performed with the CMS detector~\cite{B2G-17-009}. Again, a width of up to 30\% of the VLQ mass is considered. The search is performed in the fully hadronic final state with a boosted Higgs boson and a forward jet. A Higgs candidate is considered as an AK8 jet with a mass compatible with the Higgs boson and with 2 b-tagged subjets. High jet activity is required in the selected events. The vector-like B mass $m_\mathrm{B}$ is reconstructed from the Higgs candidate and the leading b-tagged jet. The dominant QCD multijet background is estimated from data using an ABCD method. Finally, a maximum likelihood fit is performed on the $m_\mathrm{B}$ variable, and single B production cross sections are excluded above $[0.07,1.28]$~pb, as a function of the B mass. 

Pair and singly produced vector-like Q quarks that couple to light generations and decay to W, Z, or Higgs bosons were searched for by the CMS Collaboration~\cite{1708.02510}. Two analyses were combined. The first one is an inclusive analysis that defines two event categories targeting the single production of a Q quark decaying to a W or a Z boson, and four categories optimized for Q pair production. The second analysis is an exclusive search that uses a kinematic fit to hypotheses of pair produced Q quarks decaying to W, Z, or Higgs bosons. In most of the event categories defined in the analyses, VLQ masses are reconstructed. Lower mass limits are derived from 400 to 1800 GeV, depending on the branching fraction and the single production cross section of the VLQ. 

\section{Searches for gauge and (pseudo)scalar resonances decaying to top quarks}

\begin{figure}[htb]
\centering
\includegraphics[width=0.43\textwidth]{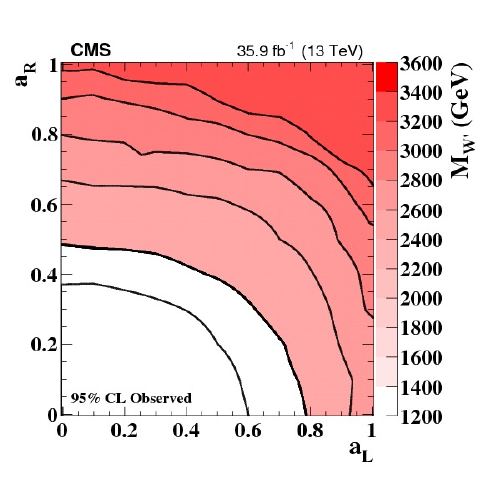}
\includegraphics[width=0.45\textwidth]{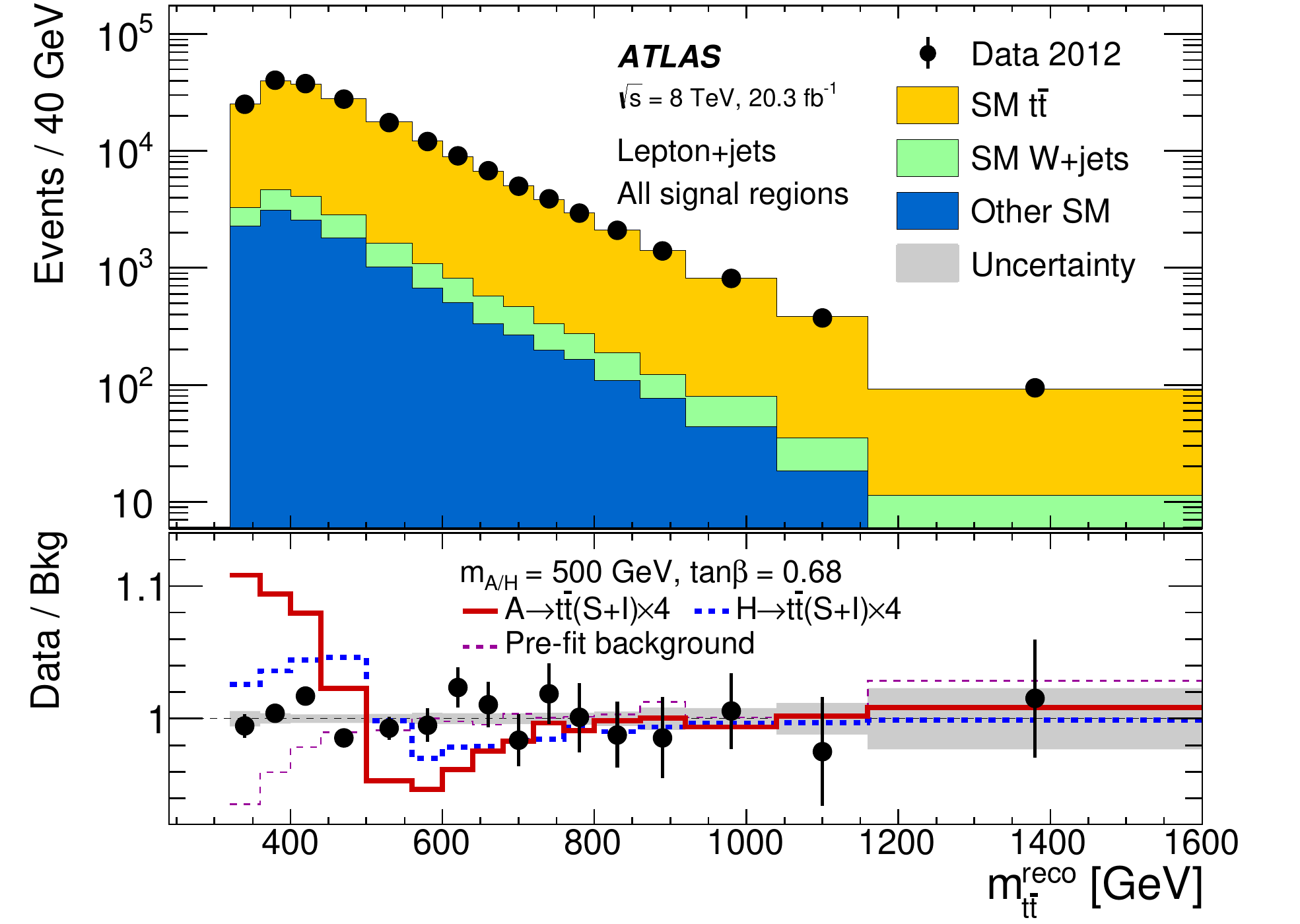}
\caption{Left: observed lower $\mathrm{W}^\prime$ mass limits as a function of right-handed and left-handed couplings~\cite{1708.08539}. Right: the presence of a heavy Higgs boson would create a peak-dip structure in the distribution of the invariant mass of the $\ttbar$ system~\cite{1707.06025}.}
\label{fig:Resonances}
\end{figure}

The CMS Collaboration performed a search for heavy $\mathrm{W}^\prime$ resonances decaying to a top and a bottom quark, in the single lepton final state~\cite{1708.08539}. The $\mathrm{W}^\prime$ mass is reconstructed from a top quark candidate and the highest-$\pt$ jet that is available after top quark candidate reconstruction. Four categories are designed based on the lepton flavor, the $\pt$ of the sum of the 2 leading jets, and the number of b-tagged jets among the leading jets. The reconstructed $\mathrm{W}^\prime$ mass distribution is used to search for a signal. As shown in Fig.~\ref{fig:Resonances} (left), mass limits are set as a function of the right-handed ($a_\mathrm{R}$) and left-handed ($a_\mathrm{L}$) $\mathrm{W}^\prime$ couplings, taking into account the quantum interference with SM single-top production. 

A search for heavy scalar (H) or pseudoscalar (A) Higgs bosons, decaying to a top quark pair, was performed with the ATLAS detector~\cite{1707.06025}. The search is performed in the single lepton channel, and the semileptonic $\ttbar$ system is reconstructed via a kinematic fit. Quantum interference effects between the SM top quark pair production and the heavy Higgs boson signal is taken into account. The interference would create a peak-dip structure in the distribution of the invariant mass of the $\ttbar$ system, as demonstrated in Fig.~\ref{fig:Resonances} (right).
The results are interpreted in a Type-II two-Higgs-doublet model. Exclusion limits are calculated in the two-dimensional plane of $\tan \beta$, defined as the ratio of the vacuum expectation values of the two Higgs doublets, and the heavy Higss boson mass, for different scenarios of the $\mathrm{H}-\mathrm{A}$ mass splitting. 

\section{Summary}

Vector-like quarks and new resonances that decay to top quarks are predicted in many models that try to solve open problems in the standard model. These hypothetical particles correspond to an extensive range of possible signatures in proton-proton collision data. Searches are performed using the CMS and ATLAS detectors for pair and single production of vector-like quarks decaying to W, Z, and Higgs bosons and third- and light-generation quarks. Other dedicated searches focus on gauge or (pseudo)scalar resonances decaying to top quarks. No evidence for new particles has been observed yet, and stringent constraints on masses, couplings and cross sections are derived.

\end{document}